\documentclass{iau} 
\usepackage{graphicx,url}

\title[H$\alpha$ imaging for BeXRBs in the SMC] %% give here short title %%
{H$\alpha$ imaging for BeXRBs in the Small Magellanic Cloud}

\author[Maravelias et al.]   %% give here short author list %%
{G. Maravelias$^1$, A. Zezas$^{2,3,4}$, V. Antoniou$^{3}$, \\ D. Hatzidimitriou$^5$, F. Haberl$^6$}

\affiliation{
$^1$Astronomick\'y \'ustav AV\v{C}R, v.v.i., Ond\v{r}ejov, Czechia, email: {\tt maravelias@asu.cas.cz} \\
$^2$Department of Physics, University of Crete, Heraklion, Greece, 
$^3$Harvard-Smithsonian Center for Astrophysics, Cambridge, USA, 
$^4$Foundation for Research and Technology-Hellas (FORTH), Heraklion, Greece, 
$^5$Department of Physics, University of Athens, Greece, 
$^6$Max-Planck-Institut f\"{u}r extraterrestrische Physik, Garching, Germany
}

\pubyear{2015}
\volume{xxx}  %% insert here IAU Symposium No.
\jname{Title of your IAU Symposium}
\editors{A.C. Editor, B.D. Editor \& C.E. Editor, eds.}
\begin{document}

\maketitle

\begin{abstract}
The Small Magellanic Cloud (SMC) hosts a large number of high-mass X-ray binaries, and in particular of Be/X-ray Binaries (BeXRBs; neutron stars orbiting OBe-type stars), offering a unique laboratory to address the effect of metalicity. One key property of their optical companion is H$\alpha$ in emission, which makes them bright sources when observed through a narrow-band H$\alpha$ filter. We performed a survey of the SMC Bar and Wing regions using wide-field cameras (WFI@MPG/ESO and MOSAIC@CTIO/Blanco) in order to identify the counterparts of the sources detected in our \textit{XMM-Newton} survey of the same area. We obtained broad-band $R$ and narrow-band H$\alpha$ photometry, and identified $\sim$10000 H$\alpha$ emission sources down to a sensitivity limit of 18.7 mag (equivalent to $\sim$B8 type Main Sequence stars). We find the fraction of OBe/OB stars to be 13\% down to this limit, and by investigating this fraction as a function of the brightness of the stars we deduce that H$\alpha$ excess peaks at the O9-B2 spectral range. Using the most up-to-date numbers of SMC BeXRBs we find their fraction over their parent population to be  $\sim0.002-0.025$ BeXRBs/OBe, a direct measurement of their formation rate.

\keywords{Magellanic Clouds, stars: early-type, stars: emission-line, Be, X-rays: binaries}
\end{abstract}

\section{Introduction}
The Small Magellanic Cloud (SMC) has been a major target for X-ray surveys due to our ability to detect sources down to non-outbursting X-ray luminosities $(L_X\sim10^{33}\,\textrm{erg s}^{-1})$ and its impressive larger number of High-Mass X-ray Binaries (HMXBs; \cite{Haberl}). However, the X-ray properties alone cannot fully characterize the nature of each source. HMXBs consist of an early-type (OB) massive star and a compact object (neutron star or black hole), which accretes matter from the massive star either through strong stellar winds and/or Roche-lobe overflow in supergiant systems or through an equatorial decretion disk in, non-supergiant, OBe stars (Be/X-ray Binaries; BeXRBs). The compact object dominates the X-ray spectrum while the companion dominates the optical spectrum. Thus, to understand the nature of BeXRBs we need to study their optical counterparts, which should be consistent with OBe stars. These are massive stars that show Balmer lines in emission, of which H$\alpha$ is typically the most prominent. Although the SMC is close enough to resolve its stellar population, we still lack the identification of the optical counterparts or their optical spectral classification for a large fraction ($\sim40\%$ of the candidate HMXBs) of the most recent census (121 candidates in total; \cite{Haberl}). To address this issue we take advantage of the fact that OBe stars display H$\alpha$ in emission, making them easily discernible from other stars in H$\alpha$ narrow-band images, and we performed a wide H$\alpha$ imaging survey of the SMC to reveal prime candidates for BeXRB optical counterparts.

\section{Observations and Data Reduction}

We used the Wide Field Imager (WFI@MPG/ESO 2.2m, La Silla, on 16/17 November, 2011) and the MOSAIC camera (@CTIO/Blanco 4m, Cerro Tololo, on 15/16 December, 2011) to observe 6 and 7 fields in the SMC, respectively. Given their large field-of-views ($\sim33'\times33'$) we covered almost the whole galaxy. Each field was observed in the $R$ broad-band (the continuum) and H$\alpha$ narrow-band filters. A dithering approach was needed to cover camera chip gaps, and the exposure time was selected to achieve a similar depth ($R\sim23$ mag) in both campaigns to allow for coverage of late  B-type stars at the distance of the SMC. Additionally, a set of spectrophotometric standards was observed to flux calibrate the results. \textsc{Theli}\footnote{ \url{http://www.astro.uni-bonn.de/~theli/}} was used to reduce and produce the final mosaics from the WFI data. For the MOSAIC data we retrieved the reduced data products from the NOAO online pipeline\footnote{\url{http://portal-nvo.noao.edu/search/query}} and then combined them using \textsc{Iraf}'s \texttt{mscred} package. We finally re-sampled (with \textsc{Swarp}\footnote{\url{http://www.astromatic.net/software/swarp}}) the mosaic images for each field, using a common center and frame size (for details see \cite{Maravelias}).  

Because of the high source density we performed PSF photometry with \textsc{Iraf}'s \texttt{daophot} by properly selecting its parameters for each field. However, we ran the source detection on the broad-band $R$ image only, as the same process in H$\alpha$ would result to many spurious sources due to the H\textsc{ii} regions of the SMC. This ($R$-selected) source list was used to perform photometry on H$\alpha$. We first screened the (flux-calibrated) \texttt{daophot} results to select stellar sources, according to their $\chi^2$ ($\sim1$) and \texttt{sharpness} ($|\textrm{sharp}|<0.5$) values. Since we were interested in OB stars we kept sources brighter than $R=18.7$ mag, which corresponds to B8 spectral-type stars at the distance of the SMC. We cross-correlated the two filters ($R$ and H$\alpha$) to identify the common sources and then with the MCPS catalog (\cite{Zaritsky}) to obtain their $V,B$ photometry. Using the locus of OB stars (\cite{Antoniou}) we selected the best OB candidate sources, for which we calculated their (H$\alpha-R$) index, its error, and SNR (following \cite{Zhao}). 

Since the $R$ filter includes the H$\alpha$ region the corresponding baseline for stars without any H$\alpha$ excess would be equal to (H$\alpha-R$)=0 mag. However, due to the differences between the two filters and the range of spectral types considered, this is not 0 (see \cite{Maravelias}). To overcome this we define the baseline (H$\alpha-R$) value for non-H$\alpha$ excess stars individually for each field based on the mode ($\langle \textrm{H}\alpha-R\rangle$) and standard deviation ($\sigma$) of the (H$\alpha-R$) distribution of all OB stars in each field. We consider as best H$\alpha$ emitting candidates the sources with: $(\textrm{H}\alpha-R) < \langle\textrm{H}\alpha-R\rangle - 5 \times \sigma$, and SNR$>5$.

\section{Results and Discussion}

\begin{figure}
  \begin{minipage}[b]{0.60\linewidth}
    \centering
    \includegraphics[width=0.95\linewidth]{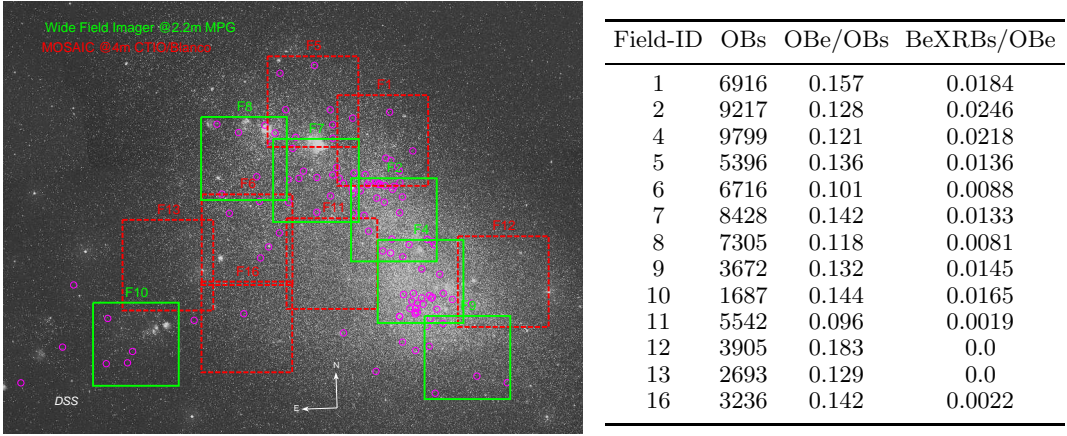} 
  \end{minipage}
  \begin{minipage}[b]{0.20\linewidth}
    \centering
\begin{tabular}[b]{ c c c c }
  \hline
  Field-ID & OBs & OBe/OBs & BeXRBs/OBe \\
  \hline 
   1 & 6916 & 0.157 & 0.0184 \\
   2 & 9217 & 0.128 & 0.0246 \\
   4 & 9799 & 0.121 & 0.0218 \\
   5 & 5396 & 0.136 & 0.0136 \\
   6 & 6716 & 0.101 & 0.0088 \\
   7 & 8428 & 0.142 & 0.0133 \\
   8 & 7305 & 0.118 & 0.0081 \\
   9 & 3672 & 0.132 & 0.0145 \\
   10 & 1687 & 0.144 & 0.0165 \\
   11 & 5542 & 0.096 & 0.0019 \\
   12 & 3905 & 0.183 & 0.0  \\
   13 & 2693 & 0.129 & 0.0  \\
   16 & 3236 & 0.142 & 0.0022 \\
  \hline  
\end{tabular}
\end{minipage}
\caption{\textit{Left:} The fields observed with the WFI (green solid boxes) and the MOSAIC (red dashed boxes) wide-field cameras, overplotted on a DSS image of the SMC. BeXRBs (taken from \cite{Haberl}) are shown as smaller purple circles. \textit{Right:} For each field (col. 1) we show the number of OB stars identified (col. 2), the fraction of H$\alpha$ emitting stars of OBe/OBs (col. 3), and the formation rate BeXRBs/OBe (col. 4). (Not including fields 3 and 14/15 due to reduction issues and shallower exposures, respectively.)}
\label{figtab}
\end{figure}

Our survey reveals 9808 H$\alpha$ emitting sources in the SMC. This is 2 to 4 times more sources from other previous surveys (1844 sources; \cite{Meyssonier}), mostly due to our deeper coverage down to $V$=18.5 mag instead of $R\sim$16.5 mag.

From our analysis we know the number of OB stars and the corresponding number of emission-line stars (i.e. OBe). This allows us to derive the OBe/OB fraction for each field. We find an average value of $\sim13.3\%$ across the SMC, consistent with previous studies (e.g. $\sim5-11\%$ from \cite{Iqbal}). This fraction is only a lower limit of the actual population of the OBe stars since their activity is a transient phenomenon and only a fraction of them is active in a certain epoch. Furthermore, if we examine the relation of this fraction with $R$ magnitude ($1\,\textrm{mag}\sim3$ spectral sub-types at the distance of the SMC), we notice a peak at $\sim$15 mag (corresponding to O9-B2) and a fast drop with magnitude (equal to later spectral types). This trend is consistent with previous observations, but we extend it to later B-type stars (from \cite{Martayan}: peak at B2 but limited to $\sim$B3). Moreover, it confirms theoretical models that predict a peak of that ratio at B3, as a result of the critical rotational velocity (\cite{Maeder}). 

Given the numbers of OBe stars and BeXRBs we derive the BeXRBs/OBe fraction in the range  $\sim0.002-0.025$, which provides us with the formation efficiency of these systems with respect to their parent population. This is a direct measurement of their formation rate, which can place constraints on stellar population synthesis models (e.g. \cite{Belczynski}). Currently, we are working on the cross-correlation of this catalog with the most recent list of candidate BeXRBs in the SMC (\cite{Haberl}) in order to identify more optical counterparts.

\vspace{3mm}
\textbf{Acknowledgements:} GA \v{C}R (14-21373S); RVO:67985815; NASA Grant NNX10AH47G; The State Scholarships Foundation of Greece (IKY); IAU travel grant.

\end{document}